\documentclass[reprint,aps,prd,nofootinbib,twocolumn,superscriptaddress,preprintnumbers]{revtex4-2}

\usepackage[T1]{fontenc}
\usepackage[usenames,dvipsnames]{xcolor}
\usepackage{aas_macros,graphicx,hyperref,mathrsfs,orcidlink,subfigure}
\hypersetup{colorlinks=true,citecolor=red,linkcolor=red,urlcolor=blue}

\usepackage{ulem}

\usepackage{graphicx,amssymb,amsmath,amsthm,amsfonts,epstopdf,epsfig,epsf,times}

\usepackage{amsmath}

\usepackage{esint}

\newcommand{\pa}[1]{\left(#1\right)}
\newcommand{\td}[1]{\tilde{#1}}

\newcommand{\df}{\mathrm{d}} 

\newcommand{\be}{\begin{equation}}
\newcommand{\ee}{\end{equation}}
\newcommand{\bea}{\setlength\arraycolsep{2pt} \begin{eqnarray}}
\newcommand{\eea}{\end{eqnarray}}
\newcommand{\nn}{\nonumber}

\def\a{\alpha}
\def\b{\beta}

\def\D{\Delta}
\def\f{\frac}
\def\g{\gamma}

\def\lm{\lambda}

\def\nn{\nonumber}

\def\p{\phi} 
\def\td{\tilde}

\def\t{\theta}
\def\T{\Theta}

\def\be{\begin{equation}}
\def\ee{\end{equation}}
\def\bag{\begin{aligned}}
\def\eag{\end{aligned}}
\def\bea{\begin{eqnarray}}
\def\eea{\end{eqnarray}}
\def\ba{\begin{array}}
\def\ea{\end{array}}

\def\bc{\begin{center}}
\def\ec{\end{center}}

\begin{document}

\title{Autocorrelation signatures in time-resolved black hole flare images: \\secondary peaks and convergence structure}

\author{Zhenyu Zhang}
\affiliation{Institute of Fundamental Physics and Quantum Technology, \& School of Physical Science and Technology, \\Ningbo University, Ningbo, Zhejiang 315211, P. R. China}
\affiliation{School of Physics, Peking University, No.5 Yiheyuan Rd, Beijing 100871, P. R. China}

\author{Yehui Hou}
\affiliation{Tsung-Dao Lee Institute, Shanghai Jiao-Tong University, Shanghai, 201210, P. R. China}
\affiliation{School of Physics, Peking University, No.5 Yiheyuan Rd, Beijing 100871, P. R. China}

\author{Minyong Guo}
\email{Contact author: minyongguo@bnu.edu.cn}
\affiliation{School of Physics and Astronomy, Beijing Normal University, Beijing 100875, P. R. China}
\affiliation{Key Laboratory of Multiscale Spin Physics, Ministry of Education, Beijing 100875, P. R. China}

\author{Yosuke Mizuno}
\email{Contact author: mizuno@sjtu.edu.cn}
\affiliation{Tsung-Dao Lee Institute, Shanghai Jiao-Tong University, Shanghai, 201210, P. R. China}
\affiliation{School of Physics \& Astronomy, Shanghai Jiao-Tong University, Shanghai, 200240, P. R. China}
\affiliation{Key Laboratory for Particle Physics, Astrophysics and Cosmology (MOE), Shanghai Key Laboratory for Particle Physics and Cosmology, Shanghai Jiao-Tong University, Shanghai, 200240, P. R. China}

\author{Bin Chen}
\email{Contact author: chenbin1@nbu.edu.cn}
\affiliation{Institute of Fundamental Physics and Quantum Technology, \& School of Physical Science and Technology, \\Ningbo University, Ningbo, Zhejiang 315211, P. R. China}
\affiliation{School of Physics, Peking University, No.5 Yiheyuan Rd, Beijing 100871, P. R. China}
\affiliation{Center for High Energy Physics, Peking University, No.5 Yiheyuan Rd, Beijing 100871, P. R. China}

\begin{abstract}

The strong gravitational field of a black hole bends light, forming multi-level images, yet extracting precise spacetime information from them remains challenging. 
In this study, we investigate how gravitational lensing leaves unique and detectable signatures in black hole movies using autocorrelation analysis. 
By examining the two-dimensional autocorrelation of a movie depicting a hotspot orbiting a Kerr black hole, as viewed by a near-axis observer, we identify a persistent secondary peak structure induced by gravitational lensing. 
Notably, these secondary peaks converge toward an approximately fixed point in the time-angle lag domain, largely independent of the orbital radius of the hotspot.
This key property suggests that combining future flare observations with precise autocorrelation analysis could effectively disentangle lensing effects from orbital dynamics, enabling direct measurement of black hole parameters. 

\end{abstract}

\maketitle

\section{Introduction}
Black holes are among the most enigmatic objects in astrophysics, offering a unique window into the nature of strong-field gravity. The Event Horizon Telescope (EHT) has provided groundbreaking observational evidence for the existence of supermassive black holes by capturing horizon-scale images of M87* and Sgr A* \citep{EventHorizonTelescope:2019dse, EventHorizonTelescope:2022wkp}. These images contain rich information about the black hole astrophysical environments, yet extracting 
the underlying spacetime geometry remains a formidable challenge. To address this, extensive efforts have been dedicated to analyzing photon rings \citep{Johnson:2019ljv, Keeble:2025gbj}, polarization patterns \citep{Gelles:2021kti, Gelles:2024tpz, Palumbo:2024czv}, and near-horizon images \citep{Chael:2021rjo, Hou:2024qqo}, among others.

A key objective of future EHT observations is the production of time-resolved ``black hole movies'' \citep{EventHorizonTelescope:2024whi}, which will capture the dynamical evolution of black hole environments with unprecedented detail. Extracting spacetime information from such time-dependent data necessitates advanced techniques. In this context, the two-point correlation function has emerged as a powerful diagnostic tool \citep{Fukumura:2007xr, Wong:2020ziu, Conroy:2023kec, Hadar:2023kau, Zhu:2023omf, Cardenas-Avendano:2024sgy, Wong:2024gph, Harikesh:2025nmt}. Strong gravitational lensing produces multiple images of an emission source, giving rise to a primary peak in the autocorrelation function, as well as subpeaks corresponding to cross-correlations between images with different orders \citep{Hadar:2020fda}. In the near future, the detectability of the observations can only reach secondary images \citep{Johnson:2024ttr}. Thus, the secondary peaks—arising from correlations between the zeroth- and first-order images—are particularly significant. However, in the presence of strong intrinsic correlations within accretion flows, these exponentially suppressed secondary peaks are often obscured by the dominant primary maximum \citep{Cardenas-Avendano:2024sgy}. The lensing-induced autocorrelation in a quasi-steady accretion flow is generally obscured by the intrinsic autocorrelation of the flow itself. Therefore, the secondary peaks of the two-dimensional angular-temporal autocorrelation have not yet been validated in previous studies based on dynamical accretion flows. 

Observations of black hole flares provide a promising avenue for overcoming this limitation. 
Flares in Sgr A*, observed in X-ray, infrared, and radio bands, are often modeled as compact hotspots \citep{Broderick:2005my, Meyer:2006fd, Trippe:2006jy, GRAVITY:2020lpa, Emami:2022ydq, Wielgus:2022heh, Vos:2022yij, Levis:2023tpb, Yfantis:2023wsp, Chen:2024jkm, Huang:2024wpj, Zhou:2024dbc}, where emission is dominated by a plasmoid with brightness exceeding the background disk by an order of magnitude \citep{Dexter:2020cuv, Baganoff:2001kw, Eckart:2005hb, Genzel:2010zy, GRAVITY:2018det, GRAVITY:2023avo}.
The predominance of plasmoids near the equatorial plane can be attributed to the magnetic topology in this region, where conditions are favorable for magnetic reconnection \citep{Ripperda:2020bpz,Ripperda:2021zpn}. The sustained presence of these plasmoids is likely associated with the formation and evolution of magnetic islands generated during the reconnection process \citep{Porth:2020txf,Ripperda:2021zpn,Najafi-Ziyazi:2023oil}. 
Crucially, the intrinsic correlation width of the hotspot scales with its size: smaller, sub-resolution hotspots yield sharper correlation peaks, thereby enhancing the visibility of lensing-induced secondary maxima.

In this paper, we present a rigorous demonstration of autocorrelation analysis as a potential tool for probing spacetime geometry through time-resolved black hole flare movies. Specifically, we perform numerical simulations of a Kerr black hole observed at low inclination angles, generating images of a compact hotspot on equatorial circular orbits. By computing the two-dimensional autocorrelation of the movie, we successfully isolate the secondary correlation peak and establish its direct correspondence to lensed photon trajectories. A particularly novel result of our analysis is the identification of a fixed point in the peak’s position, which is nearly independent of the orbital radius of the hotspot. Our theoretical framework, incorporating analytic modeling, reproduces this feature and confirms that it arises from spacetime properties. Our findings indicate that the secondary image can provide enough information for black hole spin estimation. This is remarkable, considering the fact that higher-order images are difficult to detect. We use the units $G = c = 1$ throughout, so time and length are measured in terms of the black hole mass $M$.

\section{Autocorrelation of black hole movies}
In the context of black hole imaging, the two-point correlation function of the black hole movie is defined as a six-dimensional quantity: $C_{6D}\equiv\left\langle I\left(t,\rho,\varphi\right)I\left(t^{\prime},\rho^{\prime},\varphi^{\prime}\right)\right\rangle$, where $I(t,\rho,\varphi)$ denotes the intensity at time $t$ and polar coordinates $\left(\rho,\varphi\right)$ on the observer’s screen, and the brackets $\left\langle \cdot \right\rangle$ indicate an ensemble average. 
When the system is statistically stationary and axisymmetric, the time-averaged image becomes independent of $t$ and $\varphi$, and the correlation depends only on the time lag $T = t^{\prime} - t$ and angular separation $\Phi = \varphi^{\prime} - \varphi$. 
This symmetry holds exactly for a face-on observer in Kerr spacetime and approximately at low inclinations. Under these conditions, the six-dimensional correlation reduces to a two-dimensional function \citep{Hadar:2020fda}:
\bea\label{eq:defAC2D} 
\begin{aligned}
&C_{2D}\left(T,\Phi\right)=\\
&\int\rho\df\rho\int\rho^{\prime}\df\rho^{\prime}\left\langle\Delta I\left(t,\rho,\varphi\right)\Delta I\left(t+T,\rho^{\prime},\varphi+\Phi\right)\right\rangle\,, 
\end{aligned}
\eea 
where $\Delta I\left( t,\rho,\varphi\right) = I\left(t,\rho,\varphi\right) - \left\langle I\left( t,\rho,\varphi\right)\right\rangle$ is the light intensity fluctuation, and the ensemble average can be replaced by time and angular averages: 
\bea 
\left\langle\right\rangle= \lim_{\tau \to \infty}\frac{1}{2\pi\tau}\int_{0}^{\tau}\df t\int_{0}^{2\pi}\df \varphi\,. 
\eea 
By definition, the two-dimensional autocorrelation has the reflection symmetry, i,e., $C_{2D}\left(T,\Phi\right) = C_{2D}\left(-T,-\Phi\right)$. Besides, it reaches its maximum when both the angular and time differences are zero, i.e., $C_{2D,\text{max}}\pa{T,\Phi}=C_{2D}\pa{0,0}$.

In practical implementation, we need to convert the above integrals into discrete Riemann sums. If the time interval between each frame in the black hole movie is $t_0$, and we discretize $\pa{t,\rho,\varphi}$ into $\pa{n_t,n_{\rho},n_{\varphi}}$ parts respectively, the discretized autocorrelation can be written as: 
\bea\label{eq:numAC} 
\begin{aligned}
C_{2D}&\pa{T,\Phi}\simeq N\sum_{i=1}^{n_{\rho}}\sum_{i^{\prime}=1}^{n_{\rho}}\sum_{j=1}^{n_{\varphi}}\sum_{k=1}^{n_{t}}\frac{i}{n_{\rho}}\frac{i^{\prime}}{n_{\rho}}\frac{1}{n_\varphi}\frac{1}{n_t}\\ 
&\Delta I\left(k t_0,\frac{i}{n_{\rho}},\frac{2\pi j}{n_{\varphi}}\right)\Delta I\left(kt_0+T,\frac{i^{\prime}}{n_{\rho}},\frac{2\pi j}{n_{\varphi}}+\Phi\right)\,, 
\end{aligned}
\eea 
where $N$ is an overall factor and we can define the normalized autocorrelation by  
$C\left(T,\Phi\right) \equiv C_{2D}\pa{T,\Phi}/C_{2D}\pa{0,0}$.

\medskip
\section{Dynamic imaging of hotspots}
To calculate autocorrelation, we first generate dynamic imaging of the hotspot source, producing a movie that captures the flare state.  We use the slow light treatment for dynamic imaging, where the source evolves during photon motion. This ensures that the time delay effects of the multi-level images are correctly reflected in the autocorrelation signal. We focus on the case where the hotspot undergoes geodesic circular motion. Its angular velocity satisfies 
\bea
\label{eq:omegahs}
\omega_{\text{hs}}=\frac{M^{1/2}}{r_{\text{hs}}^{3/2}+aM^{3/2}}\,,
\eea
where $r_{\text{hs}}$ is the radius of the circular motion and $a$ is the dimensionless black hole spin parameter. Such orbiting hotspot possesses a translation symmetry along $\partial_{t} + \omega_\text{hs}\partial_{\varphi}$, which means the system is invariant under $t\rightarrow t + \D t$, $\varphi \rightarrow \varphi + \omega_\text{hs} \D t$. Additionally, we assume the hotspot's emissivity decays exponentially with distance from its center \citep{GRAVITY:2020lpa}, satisfying $J\propto e^{\frac{-\boldsymbol{x}^2}{2s^2}}$, where $\boldsymbol{x}$ is the spatial distance from the emitting point to the hotspot center. Since $\boldsymbol{x}^2$ depends on both space and time, $J$ also varies accordingly, describing the dynamically evolving source system. Besides, $s$ is a physical quantity characterizing the hotspot size, and we set $s=0.3M$ in the numerical imaging scheme.

We can then use the general relativistic ray tracing and radiation transfer (GRRT) technique for imaging (see \cite{Zhang:2024lsf,Huang:2024bar} for details).
Here, we assume the hotspot is optically thin, allowing us to compute intensity by integrating the redshifted emissivity along a set of light rays. As a result, the observed intensity depends on both the screen coordinates $\pa{\rho,\varphi}$ and time $t$. Once $I(t,\rho,\varphi)$ is obtained, we can numerically compute the autocorrelation using Eqs. \eqref{eq:numAC}.

\begin{figure*}[thbp!]
	\centering
	\includegraphics[width=\textwidth]{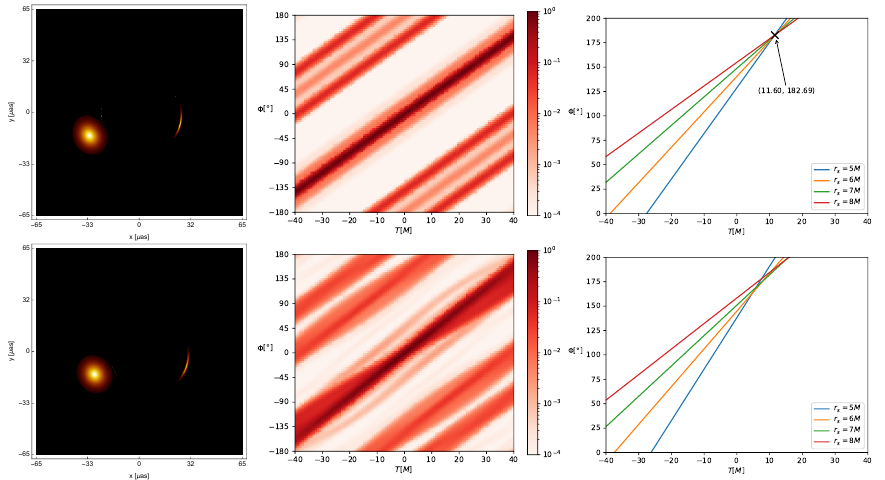}
	\centering
	\caption{\textbf{Left column}: Snapshots of the hotspot's image. The black hole spin is set to $a=0.94$, and the hotspot orbits at a radius of $r_{\text{hs}}=6M$. The axes are scaled with respect to Sgr A*. \textbf{Middle column}: Autocorrelations of the hotspot's dynamic images, where the color bar represents values on a logarithmic scale. \textbf{Right column}: Center lines of the secondary maxima. The AFP is marked with an "X". The inclinations are $\theta_o=1^\circ$ (\textbf{Top}) and $\theta_o=30^\circ$(\textbf{Bottom}), respectively. }
	\label{fig:mainresult}
\end{figure*}

\section{Numerical results}
For each parameter set used in the autocorrelation computation, we generated 100 hotspot images at a resolution of $512\times512$ pixels, sampled at intervals of one-hundredth of the orbital period. The main results, shown in Fig.\ref{fig:mainresult} for a fixed spin $a=0.94$ (with other spins in Appendix~\ref{App:DifferentSpin}), include a representative snapshot (left column) and its corresponding autocorrelation (middle column). For $\theta_\text{o}=1^\circ$, the autocorrelation shows a straight line through the origin, corresponding to the primary maximum from same-order image correlations. This line satisfies $\Phi = \omega_{\text{hs}} T$,  consistent with the pattern speed for nearly face-on observers reported in \cite{Conroy:2023kec}. In this case, the movie inherits the symmetry $\partial_{t} + \omega_\text{hs}\partial_{\varphi}$ of the hotspot, implying that the autocorrelation is symmetric under $\partial_{T} + \omega_\text{hs}\partial_{\Phi}$, as seen from Eq.~\eqref{eq:defAC2D}. 

Secondary maxima, about an order of magnitude weaker, arise from correlations between 0th- and 1st-order images. These share the same slope as the primary maximum but exhibit nonzero intercepts. 
To display an angle within $[-\pi,\pi]$ in Fig.~\ref{fig:mainresult}, we plot $\Phi$ mod $2\pi$. Due to reflection symmetry, the secondary maxima appear as two distinct branches with opposite intercepts.

Fainter parallel lines, an order of magnitude below the secondary maxima, correspond to tertiary maxima from correlations between 0th- and 2nd-order images. These features arise from gravitational lensing, reflecting the multi-image structure encoded in the black hole movie. In principle, an infinite series of such peaks may exist, tracing correlations among higher-order images, but only the tertiary maxima are visible within the limits of our numerical precision.

The top-right panel of Fig.~\ref{fig:mainresult} shows the centerlines of the secondary maxima branch with positive intercepts in the $(T,\Phi)$ plane for various $r_{\text{hs}}$\footnote{Here we show the range of $\Phi$ as $[0,200^\circ]$ instead of $[-180^\circ,180^\circ]$, because the $\Phi$ coordinate of the AFP is around $180^\circ$.}.  These lines are extracted from regions near the maxima and fitted using weighted least squares, with autocorrelation values as weights. Remarkably, all lines intersect at an approximately fixed point (AFP): $T = 11.60M$, $\Phi =182.69^\circ$. This point is obtained by averaging pairwise intersections of four fitted lines. Its existence indicates that, independent of orbital radius, the 0th- and 1st-order images remain correlated at a fixed location. This behavior likely results from the interplay between circular orbital motion and gravitational lensing in Kerr spacetime, as discussed in the next section.

EHT observations suggest a low inclination for Sgr A*, with $\theta_o \lesssim 30^\circ$ \citep{EventHorizonTelescope:2022wkp}. As inclination increases, translational symmetry breaks, yet the hotspot image remains similar to the $\theta_o = 1^\circ$ case (bottom-left panel, Fig.~\ref{fig:mainresult}). In the bottom-middle panel, secondary maxima remain discernible, though the main peak shape changes and higher-order peaks vanish. To test the persistence of the AFP at $\theta_o = 30^\circ$, we again fit four secondary-maxima centerlines using weighted least squares (bottom-right panel). While a fixed point is no longer evident, the lines still exhibit a clear tendency to converge.

\section{Semi-analytical modeling}

To provide a more explicit analysis of the hotspot's autocorrelation and the AFP, we employ a simplified model to study the hotspot's image. Since our primary focus is on the imprint of gravitational lensing on the autocorrelation, we assume that the hotspot is much smaller than the black hole, allowing us to neglect finite-size effects. For a face-on observer, the hotspot's image exhibits translational symmetry under $\partial_{t} + \omega_\text{hs}\partial_{\varphi}$, and the observed intensity can be expressed as
\bea\label{eq:thI}
I\pa{t, \rho, \varphi} = \sum_{n=0}^{\infty} F_n \delta_D\pa{\varphi-\omega_{\text{hs}} t - \varphi_n} \f{\delta_D\pa{\rho-\rho_n}}{\rho_n}\,,
\eea
where $\delta_D$ denotes the Dirac delta function, while $F_n$ and $\pa{\rho_n,\varphi_n}$ represent the flux and initial coordinates of the $n$-th order image on the screen, respectively, which remain constant for a face-on observer. Studies on strong gravitational lensing indicate that, as the order of the images increases, the area of the image undergoes exponential suppression. Consequently, the flux ratio between the images of adjacent orders approximately follows $F_{n+1}/F_n\simeq \exp{\pa{-\tilde{\gamma}_0}}$, where $\tilde{\gamma}_0$ is the Lyapunov exponent \citep{Johnson:2019ljv}. Since we are primarily interested in the secondary maxima, we retain only the contributions from the 0th and 1st order images in Eq.~\eqref{eq:thI}, yielding
\bea\label{eq:thI012}
I^{\pa{01}}\pa{t,\rho,\varphi}&\simeq &
F_0\delta_D\pa{\varphi-\omega_{\text{hs}}t}\f{\delta_D\pa{\rho-\rho_0}}{\rho_0} \nn \\
&+ & F_1\delta_D\pa{\varphi-\omega_{\text{hs}} t - \varphi_1}\f{\delta_D\pa{\rho-\rho_1}}{\rho_1}\,.
\eea
Here, we set $\varphi_0=0$, which corresponds to the initial condition where the 0th image passes through $\varphi=0$ at $t = 0$. Substituting Eq.~\eqref{eq:thI012} into the definition \eqref{eq:defAC2D}, we obtain the (unnormalized) autocorrelation:
\begin{widetext}
\bea\label{eq:thACres}
\begin{aligned}
C^{\pa{01}}(T,\Phi)=& \int\rho\df\rho \int\rho^{\prime}\df\rho^{\prime}\left\langle I^{\pa{01}}\pa{t,\rho,\varphi}I^{\pa{01}}\pa{t+T,\rho^{\prime}, \varphi+\Phi}\right\rangle \,\\
=&\frac{1}{2\pi}\left[\pa{F_0^2+F_1^2}\delta_D(\Phi-\omega_{\text{hs}}T)+F_0 F_1\delta_D(\Phi-\omega_{\text{hs}}T-\varphi_1)
+F_1 F_0\delta_D(\Phi-\omega_{\text{hs}}T+\varphi_1)\right]\,.
\end{aligned}
\eea
\end{widetext}

\begin{figure*}[thbp!]
	\centering
	\includegraphics[width=\textwidth]{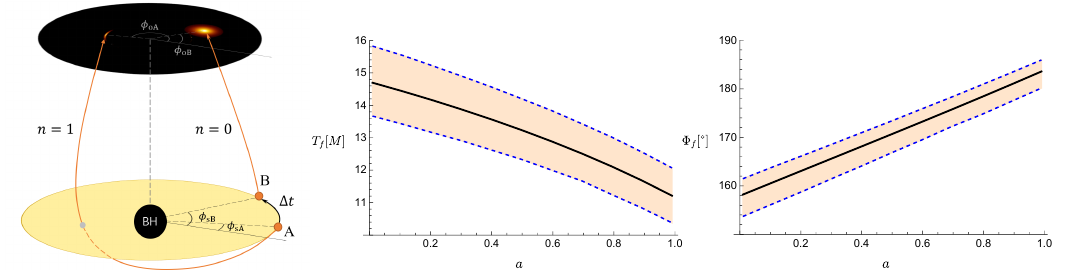}
	\centering
	\caption{\textbf{Left}: The imaging process of an orbital hotspot for a face-on observer. The axes are scaled with respect to Sgr A*. \textbf{Middle and right}: The spin dependence of the time and angular coordinates of the AFP. }
	\label{fig:2}
\end{figure*}

The first term in equation \eqref{eq:thACres} represents the primary maximum at $\Phi=\omega_{\text{hs}}T$, where $C^{\pa{01}}(T,\Phi)$ reaches its maximum value, showing excellent agreement with the numerical results in Fig.~\ref{fig:mainresult}. The second and third terms in equation \eqref{eq:thACres} indicate the presence of secondary maxima at
\bea\label{eq:secondarymax}
\Phi=\omega_{\text{hs}}T\pm\varphi_1 \,,
\eea
corresponding to two straight lines in the $(T,\Phi)$ plane with a slope of $\omega_{\text{hs}}$ and intercepts of $\pm\varphi_1$. Since $F_1/F_0\simeq \exp{\pa{-\tilde{\gamma}_0}}\simeq0.08$, the height of the secondary maxima is an order of magnitude smaller than that of the primary maximum.


The intercept of the secondary maximum is determined by the angular separation 
between the 0th- and 1st-order images of the hotspot on the observer’s screen at the same moment. The left panel of Fig.~\ref{fig:2} illustrates this for a face-on observer: as the hotspot moves from point A to B over a time $\Delta t$, light emitted from A (at $\phi = \phi_{\text{s,A}}$)
completes a half-orbit around the black hole ($n = 1$), reaching the observer with $\phi = \phi_{\text{o,A}}$ after time $t_A$.  
Meanwhile, light emitted from B (at $\phi = \phi_{\text{s,B}}$) travels directly ($n = 0$), arriving at $\phi = \phi_{\text{o,B}}$ after time $t_B$. These two rays arrive simultaneously, forming the 1st- and 0th-order images, respectively. 
Because the $n=1$ photon takes longer to reach the observer, the 0th-order image appears at a more advanced angular position, resulting in the observed offset $\varphi_1 = \phi_{\text{o,A}}-\phi_{\text{o,B}} $.
Consequently, we have
\bea\label{eq:connectAB}
\Delta t  = t_{\text{A}}-t_{\text{B}} = \f{\phi_{\text{s,A}}-\phi_{\text{s,B}}}{\omega_{\text{hs}}} \,.
\eea
By employing the gravitational lensing formula in Kerr spacetime \citep{Carter:1968rr}, we can use geodesic integrals to describe the variations in both time and angular position as light propagates from the hotspot to the observer:
\bea\label{eq:lensingAB}
\begin{aligned}
\phi_{\text{o,A}}& =  \phi_{\text{s,A}} + I^{(1)}_{\phi}(r_{\text{hs}})+\pi \,,\\
\phi_{\text{o,B}}& = \phi_{\text{s,B}} + I^{(0)}_{\phi}(r_{\text{hs}}) \,,\\
t_{\text{A}}&=I^{(1)}_{t}(r_{\text{hs}}) + a^2 G^{(1)}_{t}(r_{\text{hs}})\,,\\
t_{\text{B}}&=I^{(0)}_{t}(r_{\text{hs}}) + a^2 G^{(0)}_{t}(r_{\text{hs}})\,.
\end{aligned}
\eea
Here, $I_{t}$, $I_{\phi}$, $G_{t}$  and $G_{\phi}$ are geodesic integrals and take the forms of elliptic functions \citep{Gralla:2019ceu,Gralla:2019drh}. The superscripts indicate the order of the corresponding images. For the face-on observer, the geodesic integrals can be reduced to functions of $r_\text{hs}$ only. The details can be found in Appendix~\ref{App:geoint}. By combining equations \eqref{eq:connectAB}, \eqref{eq:lensingAB}, the intercept $\varphi_1$ of the secondary maxima can be written as
\bea\label{eq:varphia}
\begin{aligned}
\varphi_1 = & \,\, I^{(0)}_{\phi}(r_{\text{hs}}) - I^{(1)}_{\phi}(r_{\text{hs}}) + \omega_\text{hs}\left[ I^{(0)}_{t}(r_{\text{hs}}) - I^{(1)}_{t}(r_{\text{hs}})\right] \\
& + \omega_\text{hs}a^2\left[  G^{(0)}_{t}(r_{\text{hs}}) -  G^{(1)}_{t}(r_{\text{hs}})\right]  \,.
\end{aligned}
\eea
Using equations \eqref{eq:secondarymax} and \eqref{eq:varphia}, we can directly compute the secondary maxima for different black hole spins and hotspot orbital radii. We have confirmed that the results are consistent with those presented in Fig.~\ref{fig:mainresult}.

In particular, we have identified the AFP associated with the secondary maxima, showing excellent agreement with numerical results. The location of the AFP, denoted as $(T_f,\Phi_f)$, suggests a robust observational signature: irrespective of the value of $r_{\text{hs}}$,
 whenever a 0th-order image appears in a black hole movie, a 1st-order image consistently emerges near an angular displacement of $\Phi_f$	after a time delay $T_f$. Crucially, while higher-order images owe their self-similar structure to strong gravitational lensing alone, the AFP arises from a nontrivial interplay between the hotspot’s orbital motion and lensing effects. For a discussion of the impact of higher-order images, see Appendix~\ref{App:HigherOrderAC}.
 
Although the fixed point is not exact, we treat it as such in the following analytical derivation, yielding an approximate expression that demonstrates remarkable accuracy and practical utility.  The location of the AFP $(T_f,\Phi_f)$ is expected to satisfy
\bea\label{eq:fixpoint1}
\Phi_f = \omega_{\text{hs}}T_f  \pm \varphi_1 \,, \quad
0 = \left( \partial_{r_{\text{hs}}} \omega_{\text{hs}} \right) T_f \pm \partial_{r_{\text{hs}}}\varphi_1 \,,
\eea
where the second equation follows from the fixed point’s insensitivity to variations in $r_{\text{hs}}$. Accordingly, $(T_f,\Phi_f)$ is determined by
\bea\label{eq:fixpoint2}
\begin{aligned}
T_f = \mp \f{\partial_{r_{\text{hs}}}\varphi_1}{\partial_{r_{\text{hs}}} \omega_{\text{hs}}} \,, 
\quad
\Phi_f = \pm \,\varphi_1 \left( 1 - \f{\partial_{r_{\text{hs}}} \ln{\varphi_1}  }{\partial_{r_{\text{hs}}} \ln{ \omega_{\text{hs}} } } \right)  \,.
\end{aligned}
\eea
Note that equation~\eqref{eq:fixpoint2} retains a dependence on $r_\text{hs}$, indicating that $(T_f,\Phi_f)$ represents an approximate (rather than exact) fixed point. However, it still provides a novel diagnostic for constraining the black hole spin parameter, as we will discuss in section~\ref{sec:spin}.
For $a=0.94$ and $r_\text{hs}$ in the range $5M$ to $8M$, we find $\pa{T_f,\Phi_f}=\pa{11.47M\pm0.86M, 181.67^\circ\pm2.95^\circ}$, demonstrating only a weak dependence on the orbital radius. This means that if the temporal resolution of future EHT observations is larger than $1.72M$ ($\sim35\text{s}$ for Sgr A* and $\sim 16\text{h}$ for M87*), the variations in $T_f$ due to changes in $r_\text{hs}$ will be observationally indistinguishable.

The middle and right panels of Fig.~\ref{fig:2} present the evolution of the AFP as a function of the spin parameter. The left panel illustrates the variation in the time coordinate, while the right panel displays the corresponding change in the angular coordinate. The black curves represent numerical averages of the intersection points of the secondary maxima across different values of $r_\text{hs}$. The shaded region, derived from equation \eqref{eq:fixpoint2}, corresponds to the theoretical uncertainty estimate, encompassing the range where $r_\text{hs}$ varies from $5M$ to $8M$. It is evident that $T_f$ decreases monotonically with increasing spin, while $\phi_f$  increases. As $r_\text{hs}$ varies from $5M$ to $8M$, $T_f$ changes by approximately $2M$. In contrast, as the spin parameter $a$ increases from $0.01$ to $0.99$, $T_f$ varies by about $3.5M$. The angular coordinate exhibits even more pronounced variations: as $r_\text{hs}$ changes within the same range, $\phi_f$ varies by approximately $8^\circ$, whereas increasing $a$ from $0.01$ to $0.99$ results in a change of about $26^\circ$. This suggests that the observed position of $\phi_f$ could serve as a useful indicator for distinguishing between high-spin and low-spin black holes in future observations.

\section{Measuring spin by the AFP}
\label{sec:spin}

\begin{figure*}[thbp!]
	\centering
	\includegraphics[width=\textwidth]{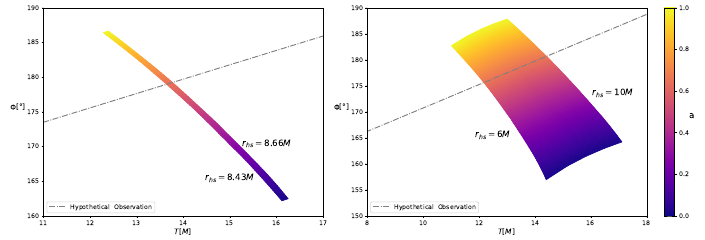}
	\centering
	\caption{The AFP shift as $a$ varies from $0$ to $1$. The gray dotted line represents the secondary maximum of a hypothetical observation with $a=0.7$ and $r_\text{hs}=8.5M$. \textbf{Left}: $r_\text{hs}$ varies from $8.43M$ to $8.66M$. \textbf{Right}: $r_\text{hs}$ varies from $6M$ to $10M$.}
	\label{fig:FPflow}
\end{figure*}
 
To further evaluate the efficacy of the AFP in characterizing black hole parameters, we present a concrete demonstration of how the AFP can be employed to estimate the black hole spin. As a specific example, consider a hotspot orbiting a black hole with spin parameter $a=0.7$ at a radius of $r_\text{hs}=8.5M$. By performing a numerical autocorrelation analysis on the simulated movie, we identify the secondary maximum, denoted by the gray dotted line in Fig.~\ref{fig:FPflow}. Using this hypothetical observational feature as the sole input, we then apply the fixed-point method to estimate the black hole spin.

For the nearly face-on case, the slope of the hypothetical observed line gives the angular velocity of the hotspot, given by equation~\eqref{eq:omegahs}. This allows us to constrain $r_\text{hs}$ to the range $8.43M$ to $8.66M$, assuming the spin parameter lies between 0 to 1. Within this interval, the trajectory of the theoretically predicted location of the AFP (given by equation~\eqref{eq:fixpoint2}) is represented by the shaded region in the left panel of Fig.~\ref{fig:FPflow}. The black hole spin can then be determined from the intersection between the hypothetical observed line and this shaded region, yielding a spin estimate of $0.699$-$0.700$. Although the AFP shifts slightly with variations in $r_\text{hs}$, we find that the resulting uncertainty in the inferred spin remains exceptionally small.

Furthermore, the fixed-point method maintains its reliability even when the estimated range of $r_\text{hs}$ is substantially relaxed. As demonstrated in the right panel of    
Fig.~\ref{fig:FPflow}, when we extend the orbital radius bounds from the precise interval $[8.43M, 8.66M]$ to a broader range of $[6M, 10M]$, the method still yields a spin estimate between $0.699$ and $0.728$, which remains remarkably accurate compared to other methods. The robustness of this approach arises from two key geometric features: (1) the consistently positive slope of the hypothetical observed line, and (2) its small acute angle relative to the equal-spin contours. Together, these properties ensure minimal variation in the inferred spin parameter across the overlapping region.

\section{summary and discussion}
By analyzing a hotspot model orbiting a Kerr black hole, we have demonstrated that gravitational lensing leaves distinct imprints on the autocorrelation of black hole movies. Numerical simulations of near-axis observations reveal a persistent secondary peak structure in the two-dimensional autocorrelation. Through theoretical modeling, we conclusively attribute this structure to lensed photon trajectories. 

Intriguingly, these secondary peaks converge to an AFP in the time-angle domain, exhibiting minimal dependence on the hotspot’s orbital radius and instead being primarily governed by the black hole’s spin and mass. This invariance provides a potential tool for constraining the spin parameter of the black hole by future time-resolved image observations. 
Future research could focus on extending the autocorrelation method to accommodate arbitrary inclination angles and different types of hotspot motion, enhancing its general applicability. Additionally, one important direction for further investigation involves examining the autocorrelation signal in the context of turbulence-driven emission variability, which could be modeled by using numerical simulations \citep{Porth:2020txf,Najafi-Ziyazi:2023oil,Jiang:2024gtk}. Such studies would provide deeper insights into the impact of turbulent dynamics on autocorrelation analyses and improve the robustness of the method under realistic astrophysical conditions.

\section*{acknowledgements}
The work is partly supported by NSFC Grant Nos. 12205013 and 12275004. YM is supported by the National Key R\&D Program of China (Grant No. 2023YFE0101200), the National Natural Science Foundation of China (Grant No. 12273022), and the Shanghai Municipality Orientation Program of Basic Research for International Scientists (Grant No. 22JC1410600).

\bibliographystyle{utphys}

\bibliography{achspt}

\clearpage

\appendix

\onecolumngrid

\section{Numerical results for different spins and radii}
\label{App:DifferentSpin}
\begin{figure}[htbp!]
	\centering
	\includegraphics[width=5.8in]{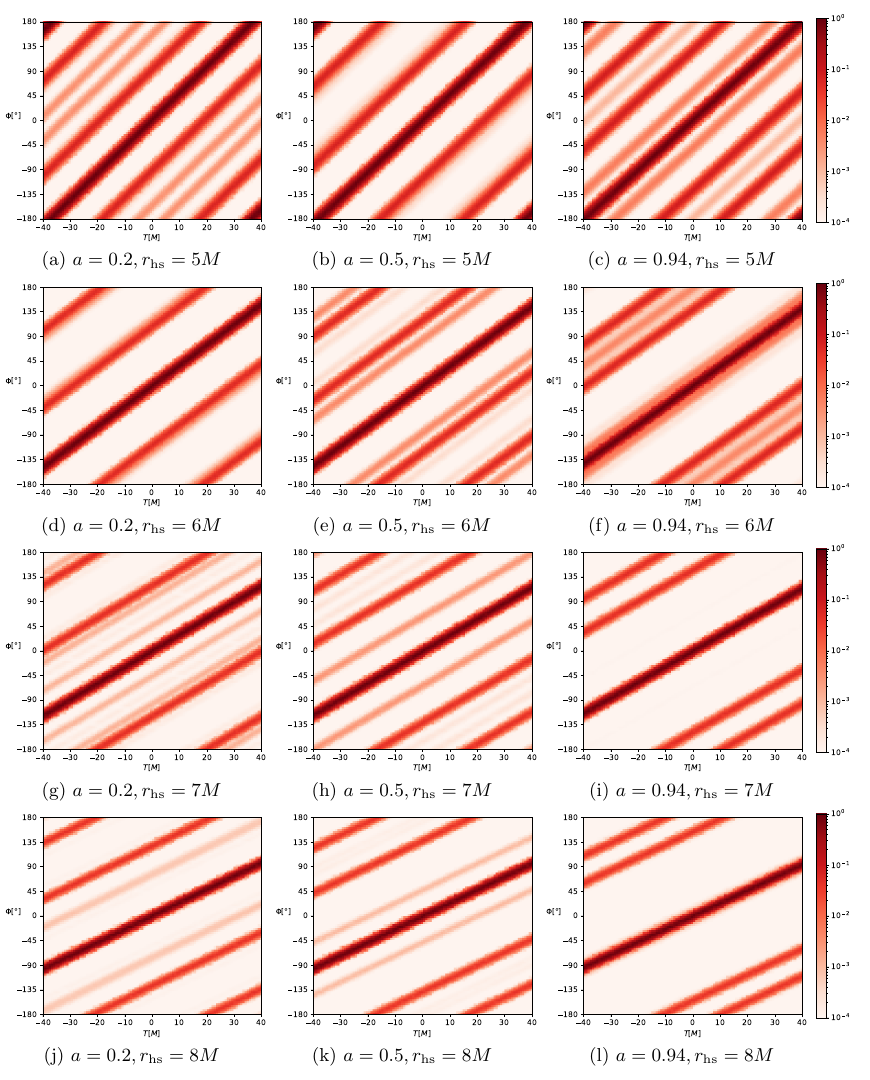}
	\centering
	\caption{Autocorrelation for different spins and radii.}
\label{fig:a02a05a094rs5678obs1}
\end{figure}

In Fig.~\ref{fig:a02a05a094rs5678obs1}, we present the numerical results for the autocorrelation function corresponding to different black hole spins and hotspot orbital radii. From left to right, each column represents a black hole spin of $a=0.2, 0.5, 0.94$, respectively. From top to bottom, each row corresponds to a hotspot orbital radius of $r_s=5M, 6M, 7M , 8M$, respectively. The observer's inclination angle is fixed at $\theta_o=1^\circ$.

\begin{figure}[htbp]
\centering
\includegraphics[width=\textwidth]{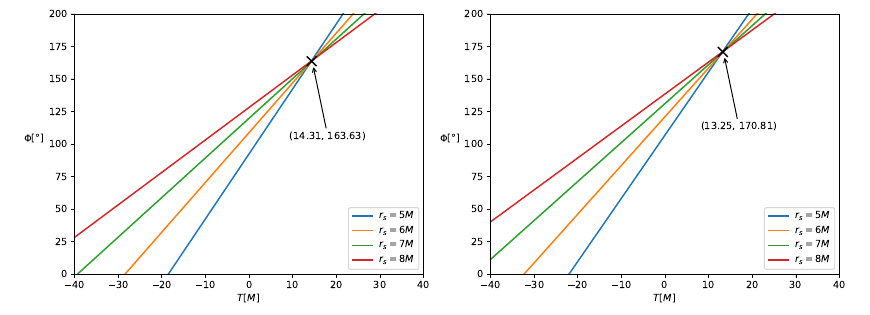}
\caption{Secondary maximum fixed points for different black hole spin cases. We fixed $\theta_o=1^\circ$. The specific coordinate values of this fixed point are obtained by finding the intersection points of four straight lines in pairs and then averaging them. \textbf{Left}: $a=0.2$. \textbf{Right}: $a=0.5$.}
\label{fig:a02a05ACFP}
\end{figure}

For varying spin values, the secondary maximum in the autocorrelation function consistently exhibits an AFP; however, its exact position shifts depending on the spin, as illustrated in Fig.~\ref{fig:a02a05ACFP}. In this figure, we again adopt an observer inclination of $\theta_o = 1^\circ$. 

\section{Geodesic integrals for photons}
\label{App:geoint}
In Kerr spacetime, the null geodesic equations governing the propagation of light between the points
$\left( t_s, r_s, \t_s, \phi_s \right)$ and $\left( t_o,r_o,\t_o,\phi_o \right)$ are given by the following expressions \citep{Carter:1968rr}:
\bea\label{eq:gequaiton}  
\begin{aligned}
&I_r = G_{\theta} \,,  \\ 
&\phi_o-\phi_s
=I_{\phi} + \lambda G_{\phi}\,,  \\ 
&t_o-t_s
= I_t + a^2 G_t \,.
\end{aligned}
\eea
Here, $I_i$ and $G_i$ are one-dimensional integrals evaluated along the geodesic with respect to $r$ and $\t$, respectively, and are defined as
\bea\label{eq:integrals}
\begin{aligned}
&I_{r}=\fint_{r_s}^{r_o}\frac{dr}{\sqrt{R(r)}}\,,\quad
G_{\t}=\fint_{\t_s}^{\t_o}\frac{d\t}{\sqrt{\T(\t)}}\,, \\
&I_{\p}=\fint_{r_s}^{r_o}\frac{a(2 M r-a\lambda)}{\Delta \sqrt{R(r)}}dr\,,\quad
G_{\p}=\fint_{\t_s}^{\t_o}\frac{\csc^2\t}{\sqrt{\T(\t)}}d\t\,, \\
&I_t=\fint_{r_s}^{r_o}\frac{r^2\Delta+2Mr(r^2 + a^2 - a\lambda)}{\Delta\sqrt{R(r)}}dr\,,\quad
G_t=\fint_{\t_s}^{\t_o}\frac{\cos^2\t}{\sqrt{\T(\t)}}d\t\,,
\end{aligned}
\eea
where $\Delta=r^2-2Mr+a^2$; $\lm$ and $\eta$ denote the energy-rescaled angular momentum and the rescaled Carter constant, respectively, commonly referred to as the impact parameters;  $R\pa{r}$ and $\Theta\pa{\theta}$ represent the effective potentials in the radial and angular directions. Their explicit expressions are given by:
\bea
\begin{aligned}
R\pa{r}&=\pa{r^2+a^2-a\lambda}^2-\Delta \left[\eta+\pa{\lambda-a}^2\right]\,,\\
\Theta\pa{\theta}&=\eta+a^2\cos^2\theta-\lambda^2\cot^2\theta\,.
\end{aligned}
\eea
The potential $R(r)$ has four roots, among which only the largest root, denoted as $r_{t}$, is relevant for photons escaping to infinity. Consequently, the path integral can be rewritten as
\bea
\fint_{r_s}^{r_o} = \int_{r_s}^{r_o} \, + \,\, 2w\int_{r_{t}}^{r_s}  \,,
\eea
where $w = 0,1$ represents the number of turning points, indicating the number of times the photon reaches $r = r_{t}$. Specifically, photons with $w = 0$ correspond to emergent rays, while those with $w = 1$ correspond to reflected rays. The motion in the $\t$ direction exhibits an oscillatory behavior between an upper and a lower bound, denoted as $\t_{\pm}$. Accordingly, the path integral takes the form of
\bea
\fint_{\t_s}^{\t_o} =  2n\int_{\theta_-}^{\t_+}
\pm_s\int_{\t_s}^{\pi/2} \mp_o\int_{\pi/2}^{\t_o} \,,
\eea
where $n = 0,1,2 ...$ represents the number of turning points. For photons emitted from an equatorial hotspot and reaching a distant face-on observer, the initial and final angular positions are given by $\t_s = \pi/2$, $\t_o = 0$, respectively, with the observer located at $r_o \rightarrow \infty$. In this scenario, the integer $n$ also corresponds to the number of times the photon crosses the equatorial plane. By analyzing the motion in the $\t$ direction, we find that only photons with zero angular momentum, i.e., $\lm = 0$, can reach the face-on observer. Furthermore, through a simple geometric analysis, the number of turning points in the radial direction satisfies:
\bea\label{turningpoint}
w = 
\left\{
\begin{aligned}
	&\,\, 1 \,,  \quad \,\,  r_s > \tilde{r}_0 \,\,\, \text{and} \,\,\, n > 0 \, , \\
	&\,\, 0 \, , \quad \,\,	\text{otherwise} \, , 
\end{aligned}
\right.
\eea
where $\tilde{r}_0$ denotes the radius of the photon sphere relevant to the face-on observer.  The first equaiton in \eqref{eq:gequaiton} simplifies to $I_r\left(r_s, \eta, n \right) = G_{\t}\left(\eta , n\right)$, and by solving this equation, we obtain the relationship between the emission radius $r_s$ and the rescaled Carter constant $\eta$, labeled as
\bea\label{eq:rseta}
\eta  = \eta^{(n)}(r_s) \,.
\eea
Due to the geodesic integrals taking the form of elliptic functions and requiring case-by-case analysis, the expression in equation \eqref{eq:rseta} is highly complex and will not be explicitly presented here. However, for $n = 0$ and sufficiently large $r_s$, an approximate formula is available \citep{Gelles:2021kti}: 
\bea
\sqrt{\eta + a^2} \,\,\approx\, \f{r_s}{M}+1+\frac{a^2-M^2}{2Mr_s}+\frac{50M^2-2a^2-15M^2\pi}{4r_s^2}+\mathcal{O}(M^3/r_s^3)\,.
\eea
Although derived under the limit $r_s \gg M$, the above equation remains a good approximation even at the horizon scale, as demonstrated in Fig. 5 of \citep{Gralla:2019drh}. Utilizing equation \eqref{eq:rseta}, the geodesic equations for the $\phi$ and $t$ directions simplify to
\bea\label{eq:phitfaceon}
\begin{aligned}
&\phi_o - \phi_s = I_{\phi}\left(r_s, \eta^{(n)}(r_s) \right) + n\,\pi \,, \\
&t_o - t_s = I_{t}\left(r_s, \eta^{(n)}(r_s) \right) + a^2 G_{t}\left(r_s, \eta^{(n)}(r_s) \right)\,.
\end{aligned}
\eea
Clearly, the variations in both time and azimuthal angle depend solely on $n$ and $r_s$. In this study, we solve equation \eqref{eq:phitfaceon} numerically. Additionally, for $n = 0$, the change in the azimuthal angle can be approximated by the following piecewise function \citep{Chen:2024jkm}:
\bea\label{appphi}
\phi_o - \phi_s \approx
\left\{
\begin{aligned}
	&\,\, \frac{2Ma}{r_s^2}\left[1-\frac{\td{\b}}{(\frac{r_s}{r_H}-1)^{\td{\a}}}\ln{(\frac{r_s}{r_H}-1)}\right] \,,  \quad \,\, r_H < r_s\textless r_{\text{ms}}\, , \\
	&\,\, \frac{2Ma}{r_s^2}\left[1+\frac{\td{\g}}{(\frac{r_s}{r_H}-1)}\right] \, , \quad \,\,	r_s\geq r_{\text{ms}}\, , 
\end{aligned}
\right.
\eea
where $r_H$ denotes the radius of the event horizon, $r_{\text{ms}}$ is the radius of the prograde innermost stable circular orbit (ISCO), and $\td{\a}, \td{\b}, \td{\g}$ are three functions of $a_{\star} = a/M$,
\bea
\left\{
\begin{aligned}
	\td{\a}&=0.035(1-a_{\star})+\frac{0.0059}{(1-a_{\star})^{0.4577}}+0.1163 \,, \\
	\td{\b}&=0.2093\arctan{(a_{\star}^{12})}+0.3467 \,, \\
	\td{\g}&=0.07815\arctan{(a_{\star}^3)}+0.0983\,.
\end{aligned}
\right.
\eea
The function \eqref{appphi} works well for a large parameter range of $a_{\star}$, as is discussed in \citep{Chen:2024jkm}.

\section{Autocorrelation of higher-order images}
\label{App:HigherOrderAC}
If we consider the effect of the second-order image, we have:
\bea
\begin{aligned}
C_{\text{th}}^{\pa{012}}&(T,\Phi)
=\pa{F_0^2+F_1^2+F_2^2}\delta_D\pa{\Phi-\omega_{\text{hs}}T}\,\\
+&F_0 F_1\left[\delta_D\pa{\Phi-\omega_{\text{hs}}T-\varphi_1}+\delta_D\pa{\Phi-\omega_{\text{hs}}T+\varphi_1}\right]\,\\
+&F_1 F_2\left[\delta_D\pa{\Phi-\omega_{\text{hs}}T+\varphi_2-\varphi_1}+\delta_D\pa{\Phi-\omega_{\text{hs}}T+\varphi_1-\varphi_2}\right]\,\\
+&F_0 F_2\left[\delta_D\pa{\Phi-\omega_{\text{hs}}T-\varphi_2}+\delta_D\pa{\Phi-\omega_{\text{hs}}T+\varphi_2}\right]\,.
\end{aligned}
\eea
We find that without using the near-photon-ring approximation, the correlations between the first-order and second-order images do not overlap at the same positions as the correlations between the zero-order and first-order images. Below we conduct the near-ring approximation and give a discussion of higher-order images. 

For higher-order images, the ratio between successive fluxes satisfies $F_{n+1}/F_n \simeq \exp{\pa{-\tilde{\gamma}_0}}$, where $\tilde{\gamma}_0$ is the Lyapunov exponent associated with the photon sphere at $r = \tilde{r}_0$. To facilitate calculations, we artificially extend the critical behavior of higher-order images to arbitrary orders. However, the resulting error in the final outcome is predominantly due to the contributions from the 0th and 1st images. Under this approximation, the intensity can be expressed as
\bea\label{eq:Ihigher}
I\pa{t, \rho, \varphi} \simeq F_0 \sum_{n=0}^{\infty} e^{-n\tilde{\gamma}_0} \delta_D\pa{\varphi-\omega_{\text{hs}} t - \varphi_n} \f{\delta_D\pa{\rho-\rho_n}}{\rho_n} \,,
\eea
and the (unnormalized) autocorrelation function is given by
\bea\label{eq:Chigher}
\begin{aligned}
C(T,\Phi)=& \int\rho\df\rho \int\rho^{\prime}\df\rho^{\prime}\left\langle I\pa{t,\rho,\varphi}I\pa{t+T,\rho^{\prime},\varphi+\Phi}\right\rangle \\
\simeq & F_0^2  \sum_{n,m} e^{-(n+m)\tilde{\gamma}_0} \lim_{\tau \to \infty}\frac{1}{2\pi\tau}\int_0^{\tau}\df t\int_0^{2\pi}\df \varphi \, \delta_D\pa{\varphi-\omega_{\text{hs}} t + \varphi_n}\delta_D\pa{\varphi+\Phi-\omega_{\text{hs}}\pa{t+T}  + \varphi_m } \\
=& \f{F_0^2}{2\pi}  \sum_{n,m} e^{-(n+m)\tilde{\gamma}_0} \delta_D\pa{\varphi_m -\varphi_n +\Phi-\omega_{\text{hs}}T} \,.
\end{aligned}
\eea
From the expression of the intensity \eqref{eq:Ihigher}, we see that $\varphi_n$ represents the azimuthal angle of the $n$-th photons on the observer's screen at $t_o = 0$. According to the lensing effect for higher-order images, the relationship between the emission and arrival positions of photons is given by \citep{Gralla:2019drh}
\bea
\begin{aligned}
&\varphi_n - \phi^{(n)}_s \simeq \left( n + \f{1}{2} \right)\tilde{\delta}_0 + D \,, \\
&t_o - t^{(n)}_s \simeq \left( n + \f{1}{2} \right)\tilde{\tau}_0  + H \,, 
\end{aligned}
\eea
where $\tilde{\delta}_0$ and $\tilde{\tau}_0$ are the critical parameters in $t$ and $\phi$ directions, respectively, associated with the photon sphere at $r = \tilde{r}_0$. The quantities $D$ and $H$ are functions of the emission radius. Therefore, we can obtain
\bea\label{eq:phithigher}
\begin{aligned}
\varphi_n - \varphi_m \simeq  \omega_{\text{hs}} (  t^{(n)}_s -  t^{(m)}_s ) + \left(n - m\right)\tilde{\delta}_0 \\
 = \omega_{\text{hs}} \left(m - n\right) \tilde{\tau}_0  + \left(n - m\right)\tilde{\delta}_0  \,, 
\end{aligned}
\eea
where we have used the relation $\phi^{(n)}_s -\phi^{(m)}_s = \omega_{\text{hs}} (  t^{(n)}_s -  t^{(m)}_s )$ for an emission source following a circular orbit. By utilizing equation \eqref{eq:phithigher}, we can rewrite the autocorrelation function as
\bea\label{eq:Chigher2}
\begin{aligned}
C(T,\Phi) \simeq &\f{F_0^2}{2\pi}  \sum_{n,m} e^{-(n+m)\tilde{\gamma}_0} \delta_D\pa{ \left(m - n\right)(\tilde{\delta}_0 -  \omega_{\text{hs}} \tilde{\tau}_0) + \Phi-\omega_{\text{hs}}T} \\ 
 =  &\f{F_0^2}{2\pi} \sum_{n = m} e^{-2n\tilde{\gamma}_0} \delta_D\pa{ \Phi-\omega_{\text{hs}}T} \\
& +
\f{F_0^2}{2\pi} \sum_{n > m}  e^{-(n+m)\tilde{\gamma}_0} \delta_D\pa{ \left(m - n\right)(\tilde{\delta}_0 -  \omega_{\text{hs}} \tilde{\tau}_0) + \Phi-\omega_{\text{hs}}T } \\
& +
\f{F_0^2}{2\pi} \sum_{n > m}  e^{-(n+m)\tilde{\gamma}_0} \delta_D\pa{ \left(n - m\right)(\tilde{\delta}_0 -  \omega_{\text{hs}} \tilde{\tau}_0)+ \Phi-\omega_{\text{hs}}T }
 \,,
\end{aligned}
\eea
and by defining $q = |n - m|$, we finally obtain the following expression:
\bea
\begin{aligned}
C(T,\Phi) \simeq \f{1}{2\pi}\f{F_0^2}{(1- e^{-2\tilde{\gamma}_0} )} \, \sum_{-\infty}^{\infty}  e^{-|q|\tilde{\gamma}_0} \delta_D\pa{ \Phi-\omega_{\text{hs}}T -q(\tilde{\delta}_0 -  \omega_{\text{hs}} \tilde{\tau}_0)} \,.
\end{aligned}
\eea
In the above expression, the term with $q = 0$ corresponds to the primary maximum, arising from the overall superposition of correlations between images of the same order. The peaks suppressed by $\exp{(-|q|\tilde{\gamma}_0)}$ originate from the correlation between images whose order differs by $q$. In the $(T,\Phi)$ plane, the intercepts of these peaks of the autocorrelation function are given by $q(\tilde{\delta}_0 -  \omega_{\text{hs}} \tilde{\tau}_0)$, indicating that for correlations between higher-order images, the intercept is jointly determined by the critical parameters and the angular velocity of the hotspot. Given that the critical parameters are approximately $\tilde{\delta}_0 \simeq \pi$, $\tilde{\tau}_0 \simeq 3\sqrt{3}M\pi$ \citep{Gralla:2019drh}, for an angular velocity satisfying $\omega_{\text{hs}} \ll (3\sqrt{3}M)^{-1}$, the peak intercepts are predominantly governed by $\tilde{\delta}_0$. In the case of Keplerian motion, this condition holds when $r_{\text{hs}} \gg 3M$.

\end{document}